\documentclass[a4paper,10pt,twoside]{cpc-hepnp}
\usepackage{CJK,upgreek,fancyhdr}
\usepackage{multicol}
\usepackage{graphicx}
\usepackage{booktabs}
\usepackage{amssymb,bm,mathrsfs,bbm,amscd}
\usepackage[tbtags]{amsmath}
\usepackage{lastpage}
\usepackage{booktabs}

\def\bc{\begin{center}}
\def\ec{\end{center}}
\def\be{\begin{eqnarray}}
\def\ee{\end{eqnarray}}

\newcommand{\omits}[1]{}

\begin{document}
\begin{CJK*}{GB}{gbsn}



\title{\boldmath Angular distribution  coefficients of Z(W) boson produced in $e^+e^-$ collisions at $\sqrt{s}=240$ GeV}

\author{%
	Yu-Dong Wang$^{1)}$\email{wangyudong@ihep.ac.cn}%
\quad 
      Jian-Xiong Wang$^{2)}$\email{jxwang@ihep.ac.cn}%
}
\maketitle

\address{%
Institute of High Energy Physics, Chinese Academy of Sciences, Beijing 100049, China\\
School of Physics, University of Chinese Academy of Sciences, Beijing 100049, China
}

\begin{abstract}At designed CEPC, similar to hadron collider,
the angular distribution coefficients of decay lepton pair from produced Z(W) boson in $e^+ e^-$ collisions are predicted.
Their dependence on $cos\theta_Z$($cos\theta_W$) are presented in four different polarization frame. 
Furthermore, the value of angular coefficients in different bins of $cos\theta_Z$ are presented in the C-S frame.
In comparison to the case at hadron collider, better accurate measurement for $Z(W)$ is expected since there exists less background 
and $W$ could be reconstructed from it's leptonic decay channel. 
This work supply a way to precise test the electroweak production mechanism or some effect induced from new physics
in the future measurements at the CEPC.
\end{abstract}




\begin{multicols}{2}

\section{Introduction}
The Drell-Yan process, firstly proposed in ~ref.\cite{drell-yan},  which study the angular distribution of lepton pair from vector boson decay which produced in the hadron-hadron collisions.
It is predicted in the simplest case  that  the differential cross section  is  proportional to $1+cos^2\theta$ at lowest order when the vector boson is virtual photon.
With the emission and absorption of  partons with large transverse momentum, there is  a factor of 2 enhancement  to the total cross section\cite{1982}, and the angular distribution become more general\cite{lam-tung,E.Mirkes:1994,E.Mirkes:1995}.
Through   measurements of the angular distribution coefficients of  final-state lepton,  many theoretical works such as the violation of Lam-Tung relation\cite{E.Mirkes:1994},  the forward-backward asymmetry of lepton pair productions\cite{AFB} were studied. 

The Drell-Yan type processes provide a powerful method to understand the production mechanism of gauge boson and  explore the  new physics.
In 1983,  $W$ and Z boson are discovered \cite{G. Arnison,G. Arnison1}, and some following measurements was found to be consistent with the predictions of the V-A Standard Model\cite {G. Arnison2,G. Arnison3,C. Albaja}.
The measurement of angular distribution coefficients of lepton pair  in $Z/\gamma^*$ production was first reported for $p\bar{p}$ collisions at 1.96 TeV by CDF Collaboration\cite{CDF}, and the results were  found in good agreement with the predictions of QCD fixed-order perturbation theory. The measurements were also done in the CMS and ATLAS collaborations at $\sqrt{s}$=8 TeV\cite{CMS,CMS1,ATLAS,ATLAS1,ATLAS2}.
Meanwhile, many theoretical work on predictions of the inclusive Z boson production which involving emission of partons of large transverse momenta were done  \cite{T2,T4}.
 
The circular electron positron collider(CEPC)  is proposed to build in future.
It is designed that the center-of-mass (CM) energy  would  be  has maximum energy 240 GeV and a higher luminosity than the 
linear collider \cite{cepc-design}, there will be less background compared with hadron collider.
The CEPC project  aims to precise test the properties of the higgs, Z and W boson and search for new physics.
Compared to the process at hadron collider, the similar one $e^++e^-\to Z/\gamma^*(W)+X \to l^+l^- (l^-\bar{\nu}_l)+X $  is of interest and deserved to study.

In this paper we study the angular distribution coefficients of Z boson inclusive production.
In comparison to Z boson hadroproduction, the energy of Z boson is fixed at Leading-order (LO) at $e^+e^-$ collider. 
Thus for detailed study, we present these  angular distribution coefficients dependence of  $cos\theta_Z$(cos$\theta_W$),  $\theta_Z$($\theta_W$) is polar angle of Z(W) boson in laboratory frame. 

The angular differential cross section can be written as

\begin{equation}
\label{eq:eqi1}
\frac{d\sigma}{dcos\theta_Z d\Omega}=\sum_{\lambda \lambda^\prime}\frac{d\sigma_{\lambda \lambda^\prime}}{dcos\theta_Z}f_{\lambda \lambda^\prime}(\theta,\varphi)
\end{equation} 
Where  
$\theta$ and $\varphi$ are polar and azimuthal angles of the lepton in Z(W) rest frame, and $d\Omega=dcos\theta d\varphi$.
$d\sigma_{\lambda \lambda^\prime}$ and $f_{\lambda \lambda^\prime}$are production density matrix of process $e^+ +e^-\to Z/\gamma^*+X$ and decay density matrix of $Z/\gamma^*\to l^++l^-$ respectively.

The content of this paper is divided into following parts.
In Sec.2 we  present the general expression of the lepton  angular distribution of this process, and also represent the angular coefficients by the  gauge boson production density matrix elements.
In Sec.3 we calculate angular distribution coefficients numerically for total and the differential cross section in  different  polarization frame.
In Sec.4 we draw the figures of angular coefficients of Z($W^-$) production  dependence of $cos\theta_Z$ ($cos\theta_W$).
Also the calculation of coefficients of Z production processes at different bins of $cos\theta_Z$ are done.
Then the summary and conclusion is given at last Sec.


\section{The angular distribution of the  lepton pair}
For simplicity,  following discussion focus on the Z boson production and the situation is same for W boson. 
In the process that $e^+(p_1)+e^-(p_2) \to Z(p_Z)+X(p_X) \to l^+(k_1) + l^- (k_2)+ X(p_X)$($l$ is $\mu$ or $e$),
there are two planes need to define which named production plane and decay plane. In the lab frame, the first one is formed by beam direction 
and $\vec{p}_{Z}$, between which the angle  is $\theta_Z$. The other one is formed by $\vec{p}_{Z}$ and $\vec{k}_{1}$, the corresponding angle 
in Z boson rest frame is $\theta$. Finally the angle between the production and decay plane is  $\varphi$, which is invariant under 
lorentz transformation from the lab frame to the Z-rest frame.

The dilepton angular distribution is defined in the Z boson rest frame,
and we use this frame for all  following discussion.
The invariant mass window of Z boson is chosen around 91.19 GeV, the contribution from $\gamma^*$ is suppressed by a large factor and lead to a error less than  1\%.
The momentum of  Z boson, $l^-$ and  $l^+$ are expressed as
\begin{equation}
\label{eq:eq1}
\begin{split}
p_Z&=(E,0,0,0),\\
k_1&=\frac{E}{2}(1, sin \theta cos\varphi, sin\theta sin\varphi, cos\theta), \\
k_2&=\frac{E}{2}(1, -sin\theta cos\varphi, -sin\theta sin\varphi, -cos\theta),
\end{split}
\end{equation}

Where the mass of fermion (mass of $e,\mu,u,...$) is set to zero approximately.
There are four commonly used polarization frames\cite{polar-frame} correspond  different choice  $Z$-axis, 
which are recoil(helicity) frame($\vec{Z}=-\frac{\vec{p}_{1}+\vec{p}_{2}}{|\vec{p}_{1}+\vec{p}_{2}|}$),   Gottfried-Jackson frame($\vec{Z}=\frac{\vec{p}_{2}}{|\vec{p}_{2}|}$),
target frame($\vec{Z}=-\frac{\vec{p}_{1}}{|\vec{p}_{1}|}$) and
the Collins-Soper(C-S)\cite{collins-soper}($\vec{Z}\propto \frac{\vec{p}_{1}}{|\vec{p}_{1}|}+\frac{\vec{p}_{2}}{|\vec{p}_{2}|}$). 
The three-vector $\vec{p}_{1}$ and $\vec{p}_2$ used here refer to the Z boson-rest frame.
The last frame was frequently used in the measurements in hadron collision.
For example, the polarization vector in helicity frame are  expressed as
\begin{equation}
\label{eq:eq2}
\begin{split}
\epsilon_\pm=(0,\mp\frac{1}{\sqrt{2}},-\frac{i}{\sqrt{2}},0), \quad
\epsilon_0=(0,0,0,1).
\end{split}
\end{equation}

The amplitude of each channel($X_i$) in inclusive Z boson production  can be written as

\begin{equation}
\label{eq:eq2}
\begin{split}
M_i=& M_{e^+e^-\to ZX_i}^\mu \frac{-i(g_{\mu \nu}-\frac{p_{z\mu} p_{z\nu}}{m_z^2+im_z\Gamma }) }{p_z^2-m_z^2-im_z\Gamma }M_{Z\to l^+l^-}^\nu\\
=& \sum_\lambda M_{e^+e^-\to ZX_i}^\mu \frac{\epsilon_{\lambda\mu} \epsilon^*_{\lambda\nu}}{m_z\Gamma} M_{Z\to l^+l^-}^\nu\\
=&\sum_\lambda a_{\lambda}(X_i)\frac{1}{m_z \Gamma}b_\lambda
\end{split}
\end{equation}

Where $\Gamma$ is the decay width of Z boson, $a_\lambda (X_i)=M_{e^++e^-\to Z+X_i}^\mu \epsilon_{\lambda\mu}$ and $b_\lambda =\epsilon^*_{\lambda\nu}M_{Z\to l^+l^-}^\nu$.
Both $a_\lambda (X_i)$ and $b_\lambda$ are Lorentz invariant. Therefore they can be calculated in different frame. 
$a_\lambda (X_i)$  and $b_\lambda$ are calculated in the lab frame and  the Z boson rest frame respectively. 
The production and decay density matrix are defined as 
\begin{equation}
\label{eq:3}
\begin{split}
	\sigma_{\lambda \lambda_{\prime}}=&\sum_ia_\lambda (X_i) a_{\lambda^\prime}^*(X_i),\\
	D_{\lambda \lambda_{\prime}} =&\sum_{s_1,s_2}b_\lambda  b_{\lambda^\prime}^*,\\
	b_\lambda = &\bar{\mu}(k_2,s_2)(ig_v \gamma_\mu+ig_a \gamma_\mu \gamma_5)\nu(k_1,s_1) \epsilon_\lambda^*, 
\end{split}
\end{equation} 
Where the decay density matrix $D_{\lambda \lambda_{\prime}}$ is easy to obtained and the production matrix $\sigma_{\lambda \lambda_{\prime}}$ 
is discussed in appendix A.

By applying  $D_{\lambda \lambda_{\prime}}$ and $\sigma_{\lambda \lambda_{\prime}}$, 
the differential cross section  is expressed as~\cite{E.Mirkes:1994}
\begin{equation}
\begin{split}
\label{eq:eq4}
\frac{d\sigma}{d\Omega} \propto&(g_v^2+g_a^2)Sp^2(1+\lambda_{\theta}cos^2\theta \\ &+\lambda_{\varphi}sin^2\theta cos(2\varphi)+\lambda_{\theta \varphi}sin(2\theta) cos\varphi \\
&+\lambda_{\varphi}^\bot sin^2\theta sin(2\varphi)+\lambda_{\theta \varphi}^\bot sin(2\theta) sin\varphi ) \\
&+\alpha_{\theta} cos\theta+\alpha_{\theta \varphi} sin\theta cos\varphi
+\alpha_{\theta \varphi}^{ \bot } sin\theta sin\varphi)
\end{split}
\end{equation}

Where $\theta$ and $\varphi$ are the polar and azimuthal angles of dilepton in the rest frame of Z boson. The coefficients of each term is given as follows,
\begin{align} 
\label{eq:eq5}
S&=\sigma_{++}+\sigma_{--}+2\sigma_{00},&&\notag\\
\lambda_\theta&=\frac{\sigma_{++}+\sigma_{--}-2\sigma_{00}}{S},
&\lambda_\varphi&=\frac{2Re(\sigma_{-+})}{S},\notag\\
\lambda_\varphi^\bot&=\frac{-2Im(\sigma_{-+})}{S},
&\lambda_{\theta \varphi}&=\frac{\sqrt{2}Re(\sigma_{+0}-\sigma_{-0})}{S},\notag\\
\lambda_{\theta \varphi}^\bot&=\frac{\sqrt{2}Im(\sigma_{+0}+\sigma_{-0})}{S},
&\alpha_{\theta}&=\frac{-2A_l(\sigma_{++}-\sigma_{--})}{S},\notag\\
 \alpha_{\theta \varphi}&=\frac{ 2\sqrt{2}A_lRe(\sigma_{+0}+\sigma_{-0})}{S},
&\alpha_{\theta \varphi}^{\bot}&=\frac{2\sqrt{2}A_l Im(\sigma_{+0}-\sigma_{-0})}{S}.\notag \\
\end{align}

The asymmetry parameter  of fermion $ f$ is $A_f= \frac{2g_v g_a}{g^2_v+g^2_a}$, given in PDG\cite{pdg}, its value  are  $0.1515\pm 0.0019$ and $0.142 \pm 0.015$ for electron and muon respectively.
In this paper, $A_l=0.1515$ is used. 
According Eq.\eqref{eq:eq5}, there are additional three terms  $\alpha_\theta$, $\alpha_{\theta \varphi}$ and $\alpha_{\theta \varphi}^\bot$  compared to the case in which $J/\Psi$ production decay to lepton pair\cite{quarkonium} due to the presence of parity-violation coupling $g_a$.
When $g_a=0$, these terms disappear.
From above expressions, terms of $\alpha_\theta$, $\alpha_{\theta \varphi}$ and $\alpha_{\theta \varphi}^\bot$  which proportional to sin$\varphi$ or sin2$\varphi$ come from  contributions of imaginary part of density matrix elements.

As presented in appendix A, there are relations Eq.\eqref{eq:A5} for the real part in $\sigma_{\lambda \lambda^\prime}$  which are from the coupling $g_v$.
According to Eq.\eqref{eq:eq5}, this part does not contribute to the value of $\alpha_\theta$, $\alpha_{\theta \varphi}$.
Meanwhile $\frac{g_a}{g_v}=\frac{1-\sqrt{1-A_f^2}}{A_f}=0.076$, the contribution of imaginary part proportional to $g_a$ is small.
It is expected that the value of $\alpha_\theta$ and $\alpha_{\theta \varphi}$ are much smaller than other coefficients.
In particular, the following relations  are obtained, 

\begin{align}
\label{eq:eq4}
-1\le & \lambda_\theta \le 1,
&-1 \le &\lambda_\varphi \le1,\notag \\
\frac{-1 }{\sqrt{2}}\le & \lambda_{\theta \varphi}\le \frac{1}{\sqrt{2}},
&-2A_l\le  &\alpha_{\theta} \le 2A_l,\notag \\
-\sqrt{2}A_l \le & \alpha_{\theta \varphi}\le \sqrt{2}A_l,\notag 
&-1 \le &\lambda_\varphi^\bot \le1,\notag \\
\frac{-1 }{\sqrt{2}}\le & \lambda_{\theta \varphi}^\bot\le \frac{1}{\sqrt{2}},
&-\sqrt{2}A_l \le & \alpha_{\theta \varphi}^\bot\le \sqrt{2}A_l,\notag\\
\end{align}



\begin{center}

\centering 
\includegraphics[width=0.500\textwidth,height=.200\textheight,scale=2]{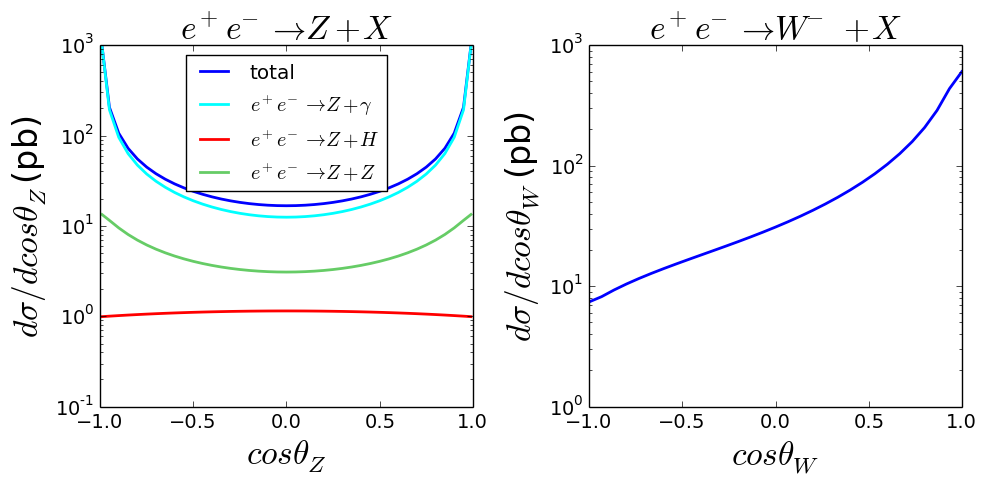}
\figcaption{\label{fig:1} The differential cross section of Z($W^-$) boson production.}
\end{center}

\newenvironment{tablehere}
{\def\@captype{table}}
{}

\section{Cross section for Z(W) boson production}
At  leading-order(LO), the production of Z boson come from following three processes
\begin{equation}
\begin{split}
\label{eq:eq9}
e^+ +e^- &\to Z + \gamma \to  l^+  +l^- + \gamma,\\
e^+ +e^- &\to Z + Z \to l^+  +l^- + Z,\\
e^+ +e^- &\to Z + H  \to l^+  +l^- + H.\\
\end{split}
\end{equation}

The production density matrices of  Z boson  in these processes are calculated by  using  package FDC\cite{fdc}, and obtain the value of angular distribution 
coefficients.

In  these three processes, the $Z \gamma$ production have a larger contribution on the cross section compared the others.
The cross section are 46.64 pb, 0.96 pb and 0.20 pb for $Z\gamma$, ZZ and ZH production channels respective.
Summed up all above  production channels, the value  of dilepton angular distribution  coefficients at Z pole are shown in the  table~\ref{tab:1}.
The total cross section   refer to the inclusive Z boson production   $e^+ +e^- \to Z + X$.

The cross section is much smaller compared with the Drell-Yan type process in hadron collisions\cite{E.Mirkes} at LO,  to obtain  accurate measurements, larger  integrated luminosity  is in need.
From the CEPC design report \cite{cepc-design}, at the CM energy $\sqrt{s}=240$ GeV,
the luminosity of the CEPC is about $3\times 10^{34}cm^{-1}s^{-1}$.
The integrated value  is about $0.8ab^{-1}/$year(it operate about 8 months each year), the total number of Z boson  produced is about $3.82\times10^7$ each year.
It is expected to run 7 years at this energy, according to the fraction of Z decay modes\cite{pdg},  the total events of  electron and muon pair  should be $1.80\times10^7$.
Also, the events of jets if we use the jets to reconstruct Z boson is about $1.87\times 10^8$.
In ATLAS\cite{ATLAS2}, the total events of lepton pair  is about $1.25\times10^7$, it is closed to the value at the CEPC.
Moreover, there is less background at the CEPC,
it is expected a good accuracy in the future CEPC experiment to measure angular distribution coefficients of inclusive Z boson production.

\end{multicols}
\begin{center}
\tabcaption{\label{tab:1} The value of  angular distribution coefficients and total cross section for Z boson productions at $\sqrt{s}=$ 240 GeV  in the Recoil frame.}
\begin{tabular}{|l| l |l |l | l | l| l | l | l | l |}
\hline
 $cos\theta_Z$&$\lambda_\theta$ &$\lambda_\varphi$ &  $\lambda_{\theta \varphi}$ &$\alpha_\theta$&$\alpha_{\theta\varphi}$& $\lambda_{\varphi}^\bot$ &$\lambda_{\theta \varphi}^\bot$ &$\alpha_{\theta \varphi}^\bot$&total cross section(pb)\\
\hline 
$cos\theta_Z>$0 & 0.937  & 0.008  &0.030  & 0.031     & -0.0003   &0&0&0&23.90\\
$cos\theta_Z<$0 &  0.937 & 0.008  & -0.030 & -0.031  & -0.003 &0&0&0&23.90\\
total& 0.937  & 0.008  &0   & 0   & -0.003  &0&0&0&47.80\\
\hline
\end{tabular}
\end{center}

\begin{center}
\tabcaption{\label{tab:2} The value of  angular distribution coefficients and total cross section for $W^-$ boson productions at $\sqrt{s}=$ 240 GeV in Recoil frame.}
\begin{tabular}{|l| l |l |l | l | l| l | l | l| l |}
\hline
 $cos\theta_W$&$\lambda_\theta$ &$\lambda_\varphi$ &  $\lambda_{\theta \varphi}$ &$\alpha_\theta$&$\alpha_{\theta\varphi}$& $\lambda_{\varphi}^\bot$ &$\lambda_{\theta \varphi}^\bot$ &$\alpha_{\theta \varphi}^\bot$&total cross section(pb)\\
\hline 
$cos\theta_W>$0  & 0.532  & -0.237  & -0.072 & 1.333  & -0.042  &0&0&0 & 96.61 \\
$cos\theta_W<$0  & 0.095 &  -0.1467 & 0.144  & -0.354 & -0.574  &0&0&0 & 9.95  \\
total& 0.486  & -0.227  & -0.048    & 1.156  & -0.099     &0&0&0& 106.64 \\
\hline
\end{tabular}
\end{center}
\begin{multicols}{2}

The value of $\lambda_\varphi^\bot$, $\lambda_{\theta \varphi}^\bot$ and $\alpha_{\theta \varphi}^\bot$ are all 0.
All these term are proportional of imaginary part of the density matrix elements in Eq.\eqref{eq:eq5}.
In the calculation, the imaginary part of density matrix is zero at LO, and these terms will  not be discussed hereafter.
	However, for Z boson production at hadron collider\cite{Wen-Chen Chang}, these coefficients are nonzero at Next-to-Leading order (NLO) corrections in QCD.

From table~\ref{tab:1}, the value of $\lambda_{\theta \varphi}$ and $\alpha_\theta$ are 0 when take full range of $cos\theta_Z$ , since they are antisymmetric in value for $cos\theta_Z> 0$ and $cos\theta_Z<0$.
Usually, there is a Lam-Tung\cite{lam-tung} relation for the coefficients $\lambda_\theta$ and $\lambda_\varphi$.  
It is presented in the Drell-Yan process that $1-\lambda_\theta =4 \lambda_\varphi$ with the vector boson being of gauge invariance condition (i.e. virtual photon)
for full final state phase space (except the dilepton) integration.
For the electroweak interaction at this situation we found that $\lambda_\theta+4\lambda_\varphi $ is about equal to 0.97, of course, it does not obey the relation
as it should be. 
From the calculation, the value of off-diagonal density matrix elements which defined in Eq.\eqref{eq:3} are about 100 times smaller compare to the diagonal matrix elements.
The conclusion is obtained from the expression of coefficients  Eq.\eqref{eq:eq5} that the value of $\lambda_\theta$ is far larger than the others, just as shown in the table~\ref{tab:1}.

The process  that $e^+ +e^- \to W^- +X \to u +\bar{d} +X$ on W boson  pole is also calculated(In the  $W^-$  production process of LO, X can only be   $W^+$).
For $W^+$ production process, the results should be symmetric with $W^-$. In table~\ref{tab:2}, the value of coefficient for  $ cos\theta_Z<0$ and $cos\theta_Z>0$ are not  symmetric or antisymmetric as Z boson production.
The contribution of the total cross section almost comes from  $cos\theta_Z>0$ region.
It is estimated  that the event number of $W^-$ boson produced at the CEPC per year at the CM energy  240 GeV  is about $8.53\times10^7$, this offer a high accuracy for W boson detection.
However, it is need to rebuild jets then to reconstruct W boson.
Comparing to hadron collider where there are a lot of source of jets due to the complicated background, 
at the CEPC there is a advantage to rebuild the jets from W boson decay with less background.
In the $W$ decay modes from PDG, the hadrons fraction is about 67.4\% and the lepton  fraction except $\tau$ is  about 21.3\%. 
By four momentum conversation, the momentum of anti-neutrino from $W^-$ decay could be obtained and then $W^-$ is reconstructed as: 
\begin{equation}
\begin{split}
\label{eq:cc}
&p_{\bar{\nu}}^2=(p_1+p_2-p_{X}-p_{e^-})^2=0,\\
&(p_{\bar{\nu}}+p_{e^-})^2=(p_1+p_2-p_{X})^2=m_W^2\\
\end{split}
\end{equation}
where $p_X$ is the sum of momentum of all the final state particles except the lepton. If $p_{\bar{\nu}}^2=0$ then it is terrified 
that this is the momentum of anti-neutrino and can be used to reconstruct the $W^-$ boson too. Finally, the total proportion of $W^-$  
which can be reconstructed through lepton and jets are about 67.4\%+67.4\%$\times21.3\%=81.8\%$.
It is deserved to measure the inclusive W boson production at the CEPC.

\section{Angular  distribution coefficients of Z(W) production processes}

We present the  differential cross section and angular distribution coefficients  dependence of the $\theta_Z$, which is the polar angle of Z boson in the lab frame.
In the experiment, usually the measurements are done in limited region.
The following  bins  in $cos\theta_Z$  which have enough events are selected  in the table~\ref{tab:3} to test the value of angular distribution coefficients.
Since the  differential cross section is symmetric in  $cos\theta_Z$, the bins in range  $-1<cos\theta_Z<0$ is just same as $0<cos\theta_Z<1$.
For the the detector at the CEPC, the
conical space with an opening angle is about 6.78 $\sim$ 8.11 degrees,
which corresponding to the value of $cos\theta_Z$ is about 0.99, i.e.
particles in the range $|cos\theta_Z|<0.99$  can be measured well.

\begin{center}
\centering 
\includegraphics[width=.50\textwidth,height=.10\textheight,scale=2]{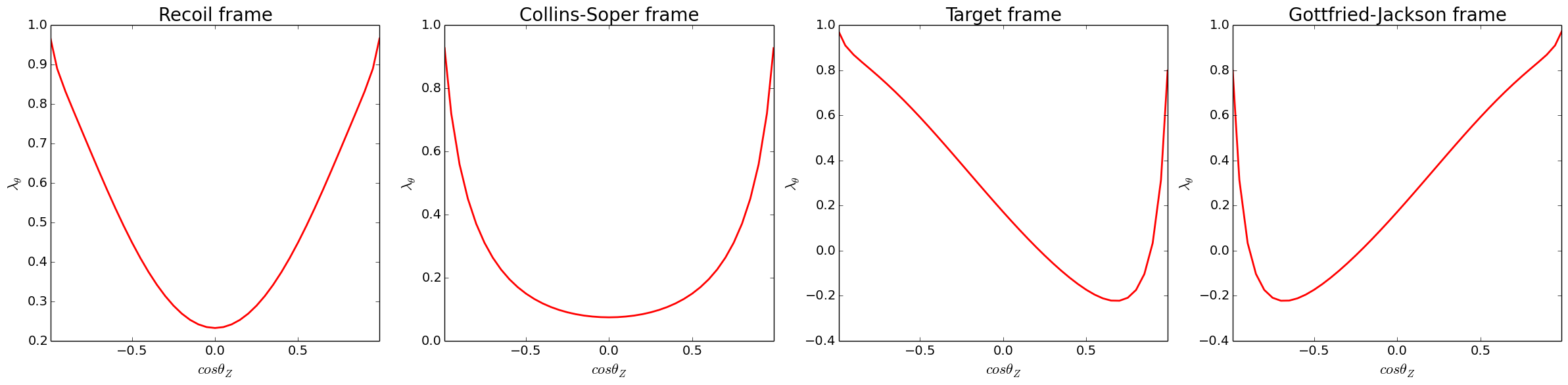}
\includegraphics[width=.50\textwidth,height=.10\textheight,scale=2]{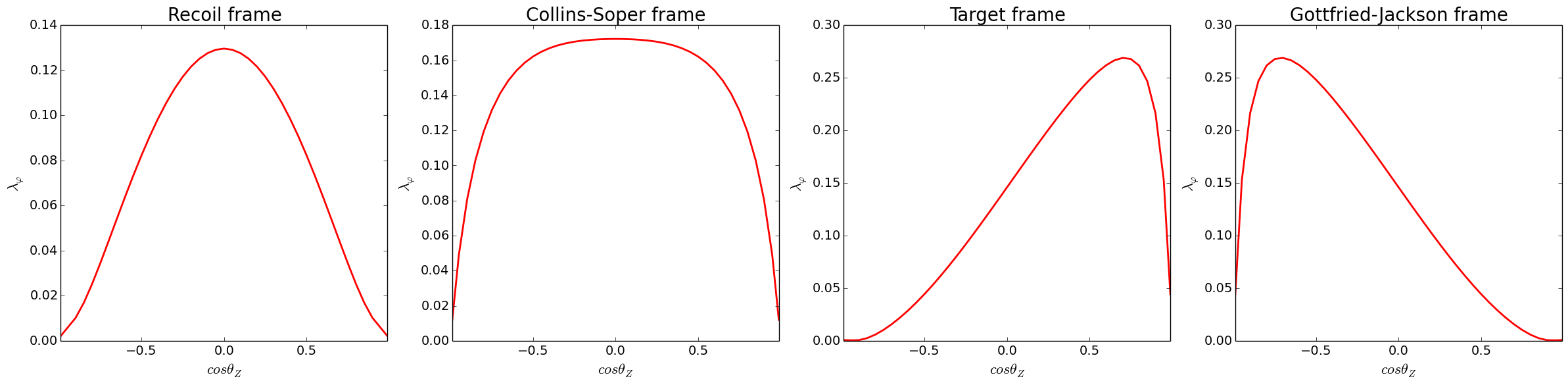}
\includegraphics[width=.50\textwidth,height=.10\textheight,scale=2]{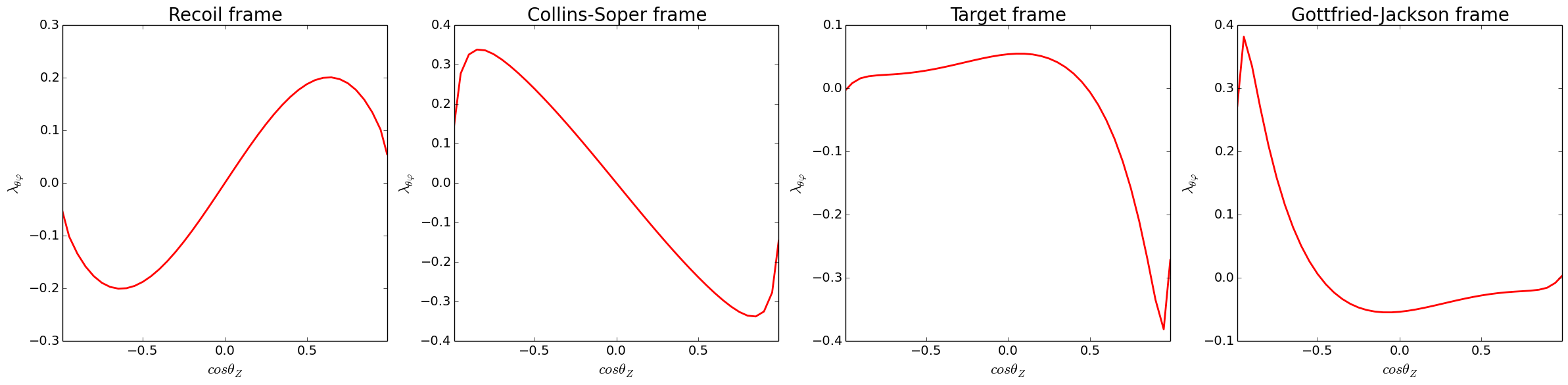}
\includegraphics[width=.50\textwidth,height=.10\textheight,scale=2]{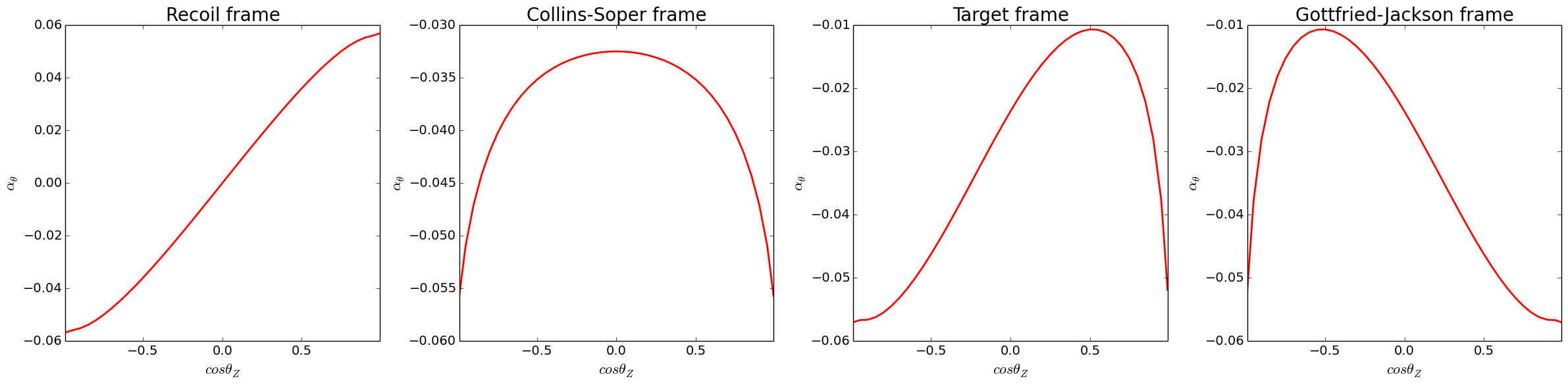}
\includegraphics[width=.50\textwidth,height=.10\textheight,scale=2]{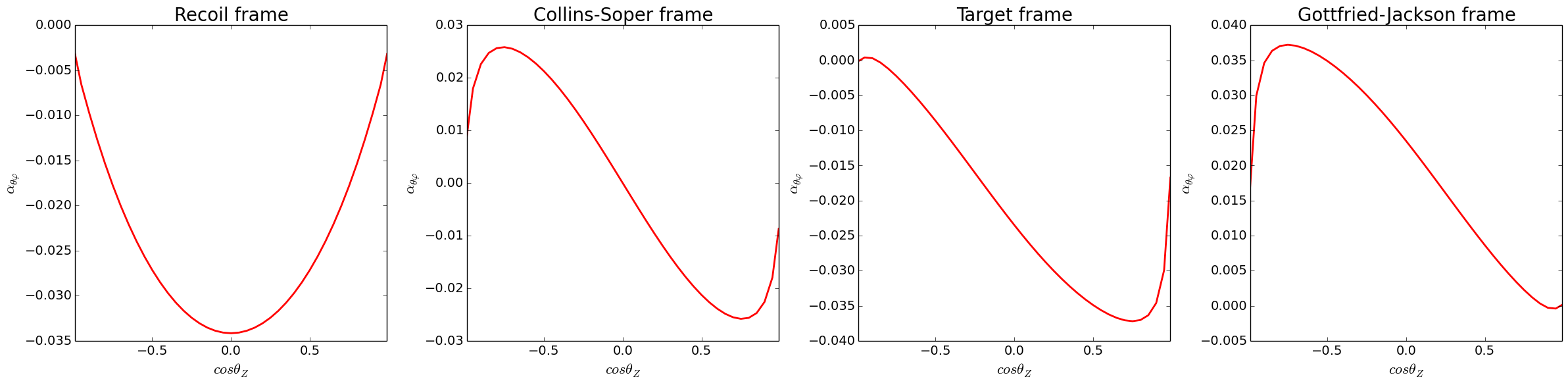}
\figcaption{\label{fig:z} Angular distribution coefficients of the inclusive Z boson production dependence on $cos\theta_Z$.}
\end{center}


 \end{multicols}
\begin{center}
\tabcaption{\label{tab:3} The  cross section for inclusive Z boson production at different bins of $cos\theta_Z$ and corresponding number of events estimated for the designed CEPC experiments.}
\begin{tabular}{|l| l |l |l | l |l| l| }
\hline
$\cos\theta_Z$  & $0-0.45 $ & $0.45-0.7 $  & $0.7-0.9 $ & $0.9-0.94 $ &$0.94-0.99 $&$0.99-1.00 $  \\
\hline
$\sigma(pb)$& 0.83  & 0.71 &1.20  &  0.54 &1.80 &20.00  \\  
\hline                                                                                                
N(1 year) & 6.6 $\times 10^5$ &5.7$\times 10^5$ &9.6$\times 10^5$  &  4.3$\times 10^5$ &1.44$\times 10^6$ &1.60$\times 10^7$  \\  
\hline                                                                                                
N(7 year) &   4.7$\times 10^6$ &4.0$\times 10^6$  & 6.7$\times 10^6$ &3.0$\times 10^6$ &1.01$\times 10^7$  &1.12$\times 10^8$\\ 
\hline
\end{tabular}
\end{center}

\begin{multicols}{2}

In the figure~\ref{fig:z}, we show the angular coefficients dependence of $cos\theta_Z$ in four different polarization frame, which are recoil frame, Collins-Soper frame, target frame and Gottfried-Jackson frame. 
In the first two frame, the angular coefficients have apparent symmetry.
For recoil frame, $\lambda_\theta$, $\lambda_\varphi$ , $\alpha_{\theta \varphi}$ are even and $\lambda_{\theta \varphi}$, $\alpha_\theta$ are odd under $cos\theta_Z \leftrightarrow -cos\theta_Z$.
For Collins-Soper frame, $\lambda_\theta$, $\lambda_\varphi$ , $\alpha_\theta$ are even and $\lambda_{\theta \varphi}$, $\alpha_{\theta \varphi}$ are odd under $cos\theta_Z \leftrightarrow -cos\theta_Z$.
Since these symmetry, as seen in the table~\ref{tab:1} that the total value of $\lambda_{\theta \varphi}$ and $\alpha_\theta$ are 0. 
The coefficients $\alpha_\theta$ and $\alpha_{\theta \varphi}$ which come from the parity-violation coupling part, are much smaller than value of  $\lambda_\theta$, $\lambda_\varphi$ and $\lambda_{\theta \varphi}$ just as the expectation discussed in Sec.2.


\begin{center}
\centering 
\includegraphics[width=.500\textwidth,height=.10\textheight,scale=2]{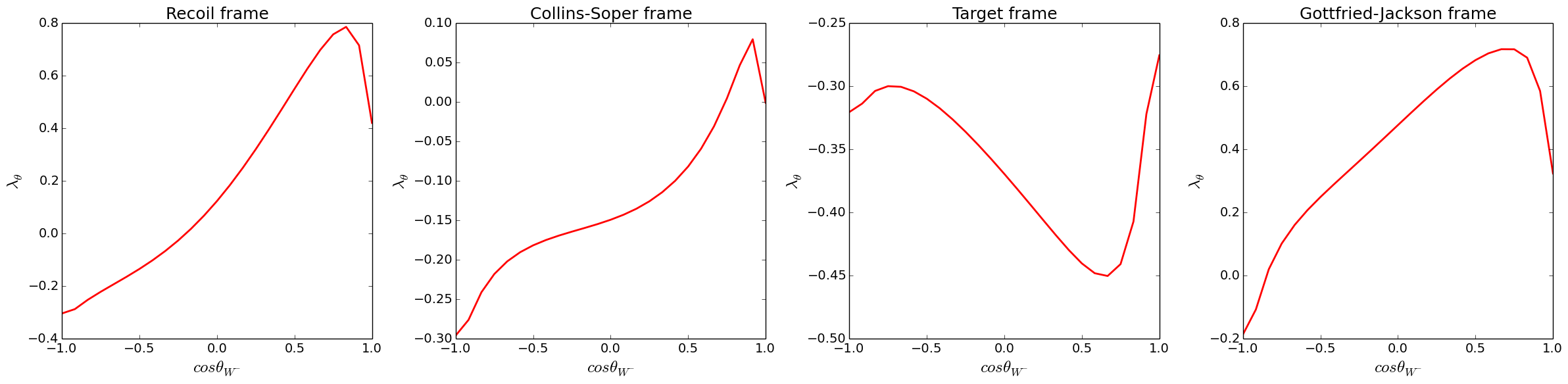}
\includegraphics[width=.500\textwidth,height=.10\textheight,scale=2]{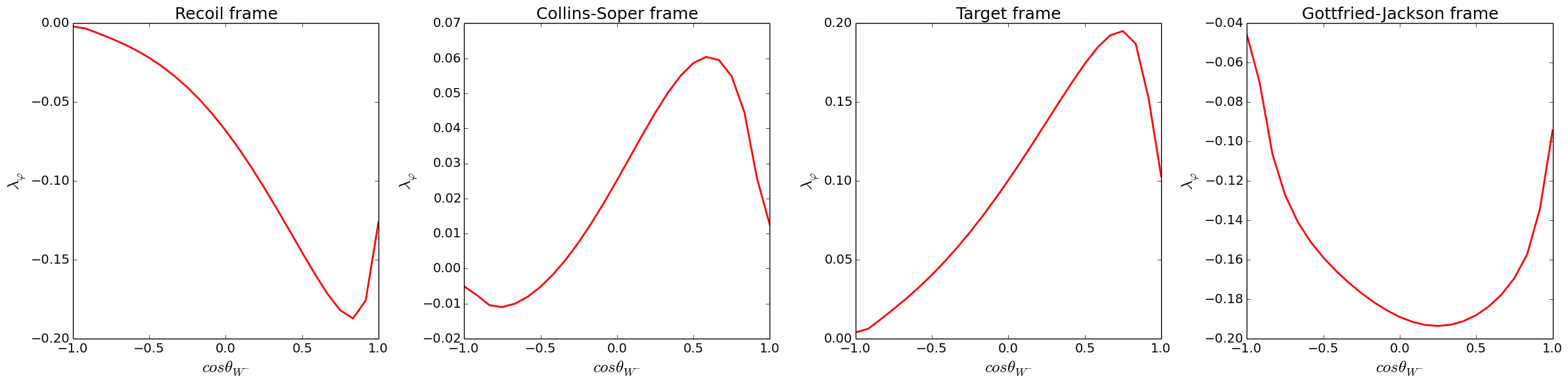}
\includegraphics[width=.500\textwidth,height=.10\textheight,scale=2]{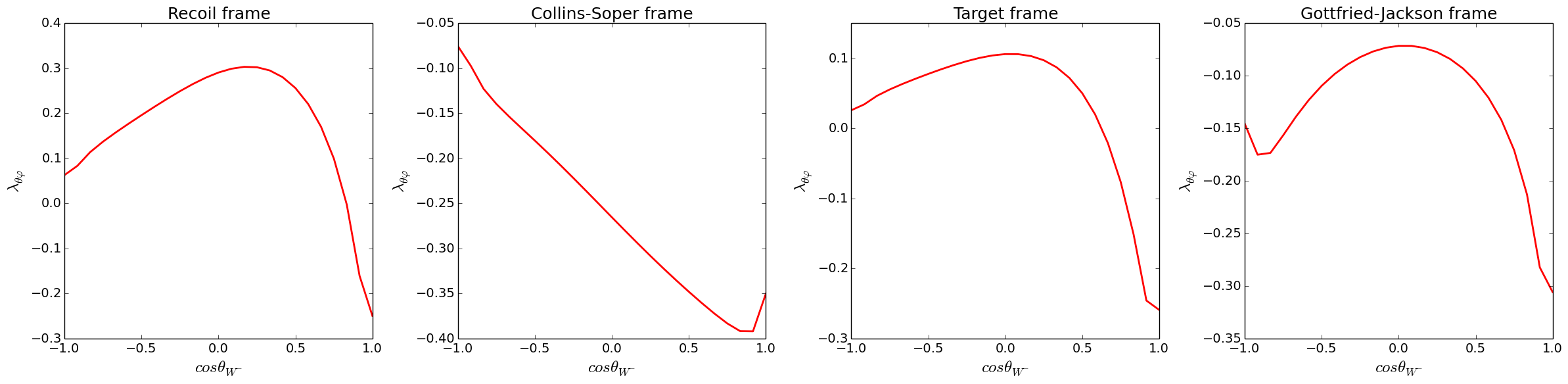}
\includegraphics[width=.500\textwidth,height=.10\textheight,scale=2]{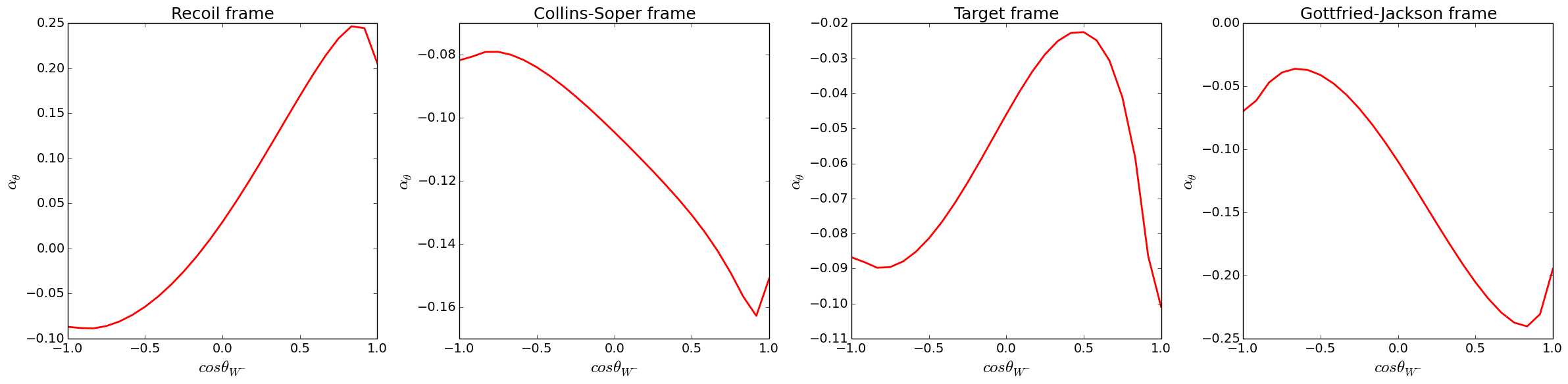}
\includegraphics[width=.500\textwidth,height=.10\textheight,scale=2]{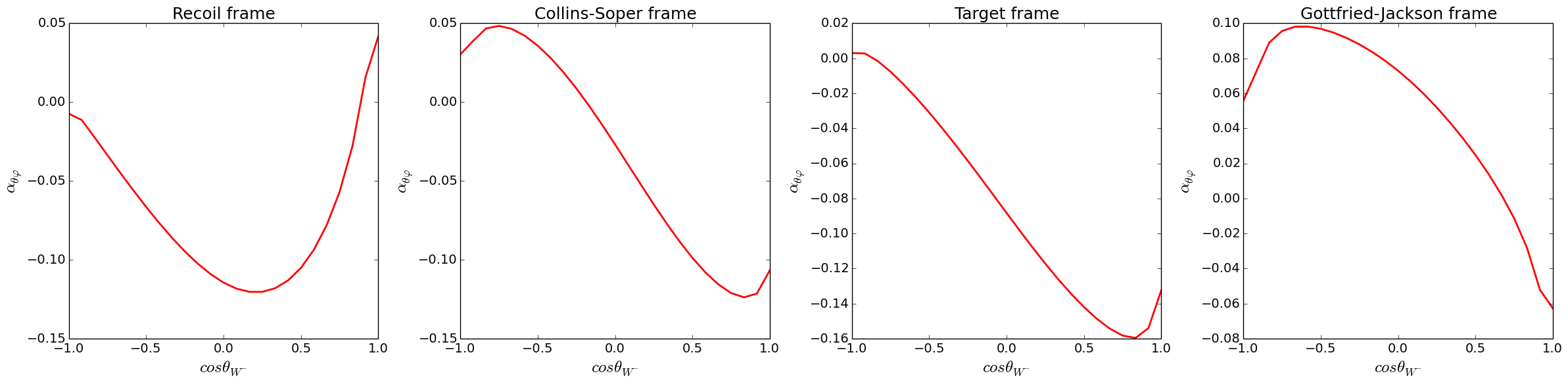}
\figcaption{\label{fig:w} Angular distribution coefficients of the inclusive $W^-$ boson production dependence on $cos\theta_W$.}
\end{center}



Then in the  figure~\ref{fig:w}, the same plots for $W^-$ productions is given.
For more discussion about $W^-$ productions, it could be refer to the process in hadron collision~\cite{W boson}.



\begin{center}
\tabcaption{\label{tab:5} The  angular distribution coefficients for Z boson production at each bin in $cos\theta_Z$ at C-S frame.}
\footnotesize
\begin{tabular}{|l| l |l |l | l |l|}
\hline
 $cos\theta_Z$&$\lambda_\theta$ &$\lambda_\varphi$ &  $\lambda_{\theta \varphi}$ &$\alpha_\theta$&$\alpha_{\theta\varphi}$\\
\hline
-$1.0-$-$0.99$   &0.996& 0 & 0.015 &-0.028 & 0\\
\hline 
-$0.99-$-$0.94$   &0.843 & 0.017 & 0.196 &-0.054 & 0.012\\
\hline    
-$0.94-$-$0.9$   &0.723 & 0.024 & 0.228 &-0.050 & 0.015\\
\hline                                                              
-$0.9-$-$0.7$    &0.480 & 0.086 & 0.291 &-0.045 & 0.022\\
\hline                                                              
-$0.7-$-$0.45$   &0.335 & 0.110 & 0.223 &-0.040 & 0.019\\ 
\hline                                                              
-$0.45-0$        &0.216 & 0.135 & 0.100 &-0.036 & 0.009\\ 
\hline                                                             
$0-0.45$         &0.216 & 0.135 &-0.100 &-0.036 &-0.009\\ 
\hline                                                              
$0.45-0.7$       &0.335 & 0.110 &-0.223 &-0.040 &-0.019\\ 
\hline                                                              
$0.7-0.9$        &0.480   & 0.086 &-0.291 &-0.045 &-0.022\\
\hline                                                             
$0.9-0.94$       &0.723 & 0.024 &-0.228 &-0.050 &-0.015\\ 
\hline 
$0.94-0.99$       &0.843 & 0.017 & 0.196 &-0.054 & 0.012\\
\hline 
$0.99-$$1.0$   &0.996& 0 & 0.015 &-0.028 & 0\\
\hline
\end{tabular}
\end{center}

\begin{center}
\tabcaption{\label{tab:6} The  angular distribution coefficients for $W^-$ boson production at each bin in $cos\theta_Z$ at C-S frame.}
\footnotesize
\begin{tabular}{|l| l |l |l | l |l|}
\hline
 $cos\theta_W$&$\lambda_\theta$ &$\lambda_\varphi$ &  $\lambda_{\theta \varphi}$ &$\alpha_\theta$&$\alpha_{\theta\varphi}$\\
\hline
$0-0.454$     & 0.328 & -0.105& 0.295 & 0.652&-0.776\\
\hline
$0.454-0.707$ & 0.635  & -0.160 & 0.209  & 1.288 &-0.597   \\
\hline
$0.707-0.891$ & 0.772  & -0.185 & 0.019  & 1.598 &-0.226  \\
\hline
$0.891-0.987$ & 0.688  & -0.171 & -0.180  & 1.592 &0.136 \\
\hline
$0.987-1$     & 0.250  & -0.098 & -0.290  & 1.206&0.356 \\
\hline
\end{tabular}
\end{center}
\bigskip

\section{Summary and conclusion}
We present the detailed definition for the theoretical calculation and experimental measurement on the lepton angular 
distribution coefficients of inclusive Z(W)  boson production needed for the designed CEPC. 
The general expression of the $cos\theta$ dependence for lepton angular distribution  coefficients in the Z  boson rest frame are 
represented by the production density matrix elements of $e^++e^-\to Z +X$, and their range is given.

From the numerical results, it is clear that the event number of lepton pair estimated at the CEPC is at same order of magnitude compared to that in the ATLAS.
The  better accurate measurements is expected since there exists less background. In comparison to case at hadron collider, the measurement for $W$ 
is of a advantage that the momentum of the missing  anti-neutrino from $W^-$ decay could be obtained and then $W^-$ is reconstructed. 
The two jets decay channels of $Z(W)$ can also be measured with less background. 
The angular distribution coefficients of  Z($W^-$) boson production dependence of $cos\theta_Z$($cos\theta_W$) is calculated in  the 
four different  polarization  frame. Furthermore, the value of angular coefficients in different bins of $cos\theta_Z$ are presented in the C-S frame.
The calculation and results in this paper supply a way to precise test the electroweak production mechanism or some effect induced from new physics 
in the future measurements at the CEPC. 

There should be study on this subject to include Monte Carlo simulation with detector and background, to include NLO electroweak correction for the 
production and $Z(W)$ boson decay, NLO QCD correction to $Z(W)$ boson decay, the correction to narrow width approximation and initial-state-radiation effect. 
all these parts is out of the scope of this work and should be addressed at future study. 
\section{Acknowledgments}

We thank Dr. Bin Gong for the discussions.
This work was supported by the National Natural Science Foundation of China with Grant No. 11475183.

\appendix
\section{The relations of the density matrix elements}
The production density matrix of the vector boson is written as

\begin{equation}
\begin{split}
\label{eq:A1}
\sigma_{\lambda \lambda^\prime}=&(M_\mu \epsilon^\mu_\lambda)(M_\nu \epsilon^\nu_\lambda)^\ast,
\end{split}
\end{equation}

where $\epsilon_\lambda$ is the polarization of the vector boson and $\lambda=+,0,-$, which is 
defined by 

\begin{equation}
\begin{split}
\label{eq:A1}
\epsilon_0=&\epsilon_z, \\
\epsilon_\pm=&\frac{1}{\sqrt{2}}(\mp\epsilon_x-i\epsilon_y), \\
\end{split}
\end{equation}

By the definition of Eq.\eqref{eq:A1} and rewrite $M_\mu$ as 
$M_\mu=M_{1\mu}+iM_{2\mu}$ that $M_{1\mu}$ and $M_{2\mu}$ 
is real(for the situation when there is no weak interaction in $M_\mu$,$M_2=0$) .

\begin{center}
\footnotesize
\begin{equation}
\begin{split}
\notag
\sigma_{+ +}=&(M_\mu \epsilon^\mu_+)(-M_\nu^\ast \epsilon^\nu_-) \\
  =&\frac{1}{2}[(M_{1\mu}M_{1\nu}+M_{2\mu}M_{2\nu})(\epsilon_x^\mu\epsilon_x^\nu+
  \epsilon_y^\mu\epsilon_y^\nu) \\
  &-(M_{2\mu}M_{1\nu}-M_{1\mu}M_{2\nu})  
  (\epsilon_y^\mu\epsilon_x^\nu - \epsilon_x^\mu\epsilon_y^\nu)],\\
\sigma_{- -}=&(M_\mu \epsilon^\mu_+)(-M_\nu^\ast \epsilon^\nu_-) \\
  =&\frac{1}{2}[(M_{1\mu}M_{1\nu} +M_{2\mu}M_{2\nu})(\epsilon_x^\mu\epsilon_x^\nu+
  \epsilon_y^\mu\epsilon_y^\nu) \\
 &+(M_{2\mu}M_{1\nu}-M_{1\mu}M_{2\nu})  
  (\epsilon_y^\mu\epsilon_x^\nu - \epsilon_x^\mu\epsilon_y^\nu)],\\
\sigma_{+ -}=&(M_\mu \epsilon^\mu_+)(-M_\nu^\ast \epsilon^\nu_+) \\
  =&-\frac{1}{2}(M_{1\mu}M_{1\nu}+M_{2\mu}M_{2\nu})[(\epsilon_x^\mu\epsilon_x^\nu-
  \epsilon_y^\mu\epsilon_y^\nu)+i(\epsilon_x^\mu\epsilon_y^\nu+\epsilon_y^\mu\epsilon_x^\nu)],\\
\sigma_{- +}=&(M_\mu \epsilon^\mu_-)(-M_\nu^\ast \epsilon^\nu_-) \\
  =&-\frac{1}{2}(M_{1\mu}M_{1\nu}+M_{2\mu}M_{2\nu})[(\epsilon_x^\mu\epsilon_x^\nu-
  \epsilon_y^\mu\epsilon_y^\nu)-i(\epsilon_x^\mu\epsilon_y^\nu+\epsilon_y^\mu\epsilon_x^\nu)],\\
\end{split}
\end{equation}
\end{center}

\begin{center}
\footnotesize
\begin{equation}
\begin{split}
\sigma_{0 +}=&(M_\mu \epsilon^\mu_0)(-M_\nu^\ast \epsilon^\nu_-) \\
   =&-\frac{1}{\sqrt{2}}\epsilon_z^\mu\{[(M_{1\mu}M_{1\nu}+M_{2\mu}M_{2\nu})
     \epsilon_x^\nu+ (M_{2\mu}M_{1\nu}-M_{1\mu}M_{2\nu})\epsilon_y^\nu] \\
      &-i[(M_{1\mu}M_{1\nu}+M_{2\mu}M_{2\nu})\epsilon_y^\nu-
      (M_{2\mu}M_{1\nu}-M_{1\mu}M_{2\nu}) \epsilon_x^\nu]\} ,\\
\sigma_{+ 0}=&(M_\mu \epsilon^\mu_+)(M_\nu^\ast \epsilon^\nu_0) \\
   =&-\frac{1}{\sqrt{2}}\epsilon_z^\mu\{[(M_{1\mu}M_{1\nu}+M_{2\mu}M_{2\nu})
     \epsilon_x^\nu+ (M_{2\mu}M_{1\nu}-M_{1\mu}M_{2\nu})\epsilon_y^\nu] \\
      &+i[(M_{1\mu}M_{1\nu}+M_{2\mu}M_{2\nu})\epsilon_y^\nu-
      (M_{2\mu}M_{1\nu}-M_{1\mu}M_{2\nu}) \epsilon_x^\nu]\} ,\\
\sigma_{0 -}=&(M_\mu \epsilon^\mu_0)(-M_\nu^\ast \epsilon^\nu_+) \\
   =&\frac{1}{\sqrt{2}}\epsilon_z^\mu\{[(M_{1\mu}M_{1\nu}+M_{2\mu}M_{2\nu})
     \epsilon_x^\nu- (M_{2\mu}M_{1\nu}-M_{1\mu}M_{2\nu})\epsilon_y^\nu] \\
      &+i[(M_{1\mu}M_{1\nu}+M_{2\mu}M_{2\nu})\epsilon_y^\nu
      +(M_{2\mu}M_{1\nu}-M_{1\mu}M_{2\nu}) \epsilon_x^\nu]\} ,\\
\sigma_{- 0}=&(M_\mu \epsilon^\mu_-)(M_\nu^\ast \epsilon^\nu_0) \\
   =&\frac{1}{\sqrt{2}}\epsilon_z^\mu\{[(M_{1\mu}M_{1\nu}+M_{2\mu}M_{2\nu})
     \epsilon_x^\nu- (M_{2\mu}M_{1\nu} -M_{1\mu}M_{2\nu})\epsilon_y^\nu] \\
      &-i[(M_{1\mu}M_{1\nu}+M_{2\mu}M_{2\nu})\epsilon_y^\nu
      +(M_{2\mu}M_{1\nu}-M_{1\mu}M_{2\nu}) \epsilon_x^\nu]\} .\\
\end{split}
\end{equation}
\end{center}

From above calculation, we see that $\sigma_{+ +}$ is not equal to $ \sigma_{- -}$, unless the situation that $M_\mu$ is real.
Besides, we obtain following relations,
\begin{equation}
\begin{split}
\label{eq:A4}
\sigma_{+ -}=&(\sigma_{- +})^\ast,\\
\sigma_{+0}=&(\sigma_{0+})^\ast,\\
\sigma_{-0}=&(\sigma_{0 -})^\ast,\\
\end{split}
\end{equation}

If $M_\mu$ is real, from Eq.\eqref{eq:A4} we could obtain relations as
\begin{equation}
\begin{split}
\label{eq:A5}
\sigma_{+ +}=&\sigma_{- -},\\
\sigma_{+ -}=&(\sigma_{- +})^\ast,\\
\sigma_{+0}=&(\sigma_{0+})^\ast=-\sigma_{0-}=-(\sigma_{-0})^\ast,\\
\end{split}
\end{equation}

\end{multicols}

\clearpage
\end{CJK*}
\end{document}